# Overcoming Ambient Drift and Negative-Bias Temperature Instability in Foundry Carbon Nanotube Transistors

Andrew C. Yu[1], Tathagata Srimani[2,3,*], Max M. Shulaker[1,4]

[1] Department of Electrical Engineering and Computer Science, Massachusetts Institute of Technology, Cambridge, MA 02139, USA

[2] Department of Electrical Engineering, Stanford University, Stanford, CA 94305, USA

[3] Department of Electrical and Computer Engineering, Carnegie Mellon University, Pittsburgh, PA 15213, USA

[4] Analog Devices, Wilmington, MA 01887, USA

*Corresponding author, contact: tsrimani@andrew.cmu.edu

**ABSTRACT**

Carbon nanotube field-effect transistors (CNFETs) are a promising candidate for back-end-of-line logic integration as a complementary path for continued electronics scaling. However, overcoming CNFET ambient drift (*i.e.*, air-stability) and reliability is underexplored. Here, we demonstrate silicon nitride encapsulation limits ambient atmosphere induced threshold voltage shift (~8× reduction of median $\Delta V_T$ over 90 days). With stabilized nitride encapsulated CNFETs, we characterize CNFET negative bias temperature instability (NBTI) with both DC and AC stress across electric field, temperature, gate oxide thickness, and stress frequency. AC pulsed operation significantly improves CNFET NBTI vs. DC operation across a wide frequency range 1 kHz to 10 MHz. 20% duty cycle AC operation at 10 MHz could extend NBTI time-to-failure by >$10^4$× vs. DC for a target $|\Delta V_T|$ tolerance ≤ 100 mV with gate bias $V_{GS,Stress}$ = -1.2 V at 125°C. This work improves our understanding of overcoming ambient drift and BTI reliability in CNFETs.

# INTRODUCTION

Back-end-of-line (BEOL) integration of transistor technologies on silicon CMOS provides a complementary scaling path along the vertical dimension[1,2]. Among the various candidates explored today[2] (e.g., carbon nanotubes, 2D Transition Metal Dichalcogenides, Oxide semiconductors), carbon nanotube (CNT) field-effect transistors (CNFETs) show great promise: (1) CNFETs overcome major process integration challenges (e.g., through new VLSI CNT purification techniques,[3] robust p- and n- type doping,[4] silicon fab-compatible CNFET contact fabrication[5]) and have undergone lab-to-fab to commercial silicon CMOS foundries [6–8]; (2) CNFET VLSI-design framework have been developed[6,9] and complex CNFET CMOS circuits, e.g., 16-bit RISC-V and SRAM arrays, have been demonstrated [6,9,10]; and (3) CNFETs are projected to achieve significant (>7×) energy delay product benefits vs. silicon FETs at extremely scaled 2-nm technology nodes[11]. However, BEOL-compatible CNFET reliability is still not well characterized, specifically the change in electrical characteristics over time and operating conditions (e.g., bias stress, temperature, target operating frequency). This presents a challenge to ensuring correct circuit operation over a device's lifetime. Especially, bias stress induced effects in threshold voltage ($V_T$) shift within BEOL CNFET circuits could significantly alter or worsen functionality and/or throughput. Prior works have studied interface traps between CNTs and gate dielectrics and their impact on CNFET hysteresis[12–17], but did not assess the reliability implications (e.g., bias-temperature instability, BTI, a key limiter in logic circuit lifetime). Other studies which have characterized BTI in CNFETs do not use silicon CMOS BEOL compatible CNFET fabrication[18,19] or use unconventional techniques to study BTI (which are therefore not used in silicon transistor BTI testing). In particular, some studies: (1) focused on long-channel/unencapsulated CNFETs with thick $SiO_2$ gate dielectric, e.g., equivalent oxide thickness, $EOT > 50$ nm, not representative of a scaled BEOL CNFET [18,20], (2) used exotic gate dielectrics and silicon-foundry incompatible processing, e.g., yttrium oxide gate dielectric, metal evaporation and lift-off, etc. [19], and/or (3) used BTI characterization techniques such as slow $I_D$-$V_{GS}$ that allows relaxation and underestimates BTI [19,21].

In this work, we address both air stability and BTI reliability of CNFETs fabricated using a silicon BEOL compatible CNFET fabrication flow on 200 mm wafers [5,7]. We characterize and improve CNFET air-stability using silicon nitride ($SiN_x$) encapsulation. With stabilized CNFETs, we characterize DC and AC negative-bias temperature instability (NBTI) for a range of gate voltage bias, stress frequency, stress duty cycle, gate dielectric equivalent oxide thickness (EOT), and temperature using similar on-the-fly BTI testing protocols used in silicon [22–24]. Our measurements and analysis show the following new insights:

(1) Silicon Nitride ($SiN_x$) encapsulation limits threshold voltage drift ($\Delta V_T$) in CNFETs, showing >8× improved median $\Delta V_T$ shift with $SiN_X$ encapsulation (median $\Delta V_T$ ~ 54 mV) vs. baseline (median $\Delta V_T$ ~ 450 mV) after 90 days in air. $SiN_x$ encapsulation thus ensures BTI measurements are not impacted by ambient $\Delta V_T$ shift during the measurement period. NBTI induced $\Delta V_T$ in $SiN_x$ encapsulated CNFETs follows similar power level dependence on gate oxide electric field $\xi_{ox}$' and stress time as observed in silicon FETs [25,26].

(2) Temperature does not increase the long-term magnitude of $\Delta V_T$ (e.g. >$10^3$ seconds), but causes the "peak" $\Delta V_T$ to occur sooner and significantly lengthens $\Delta V_T$ relaxation time.

(3) AC stress (*i.e.* pulsed gate operation) improves NBTI tolerance in $SiN_x$ encapsulated CNFETs, e.g. >$10^4$× relaxed NBTI time-to-failure using AC 20% duty cycling vs. DC for a target $|\Delta V_T|$ tolerance ≤ 100 mV at gate stress bias $V_{G,Stress}$ = -1.2 V and frequency = 10 MHz. We find AC stress NBTI $|\Delta V_T|$ is not impacted by AC frequencies from 1 kHz to 10 MHz suggesting these benefits are stable across this frequency range.

This is the first comprehensive experimental analysis of DC and AC stress induced NBTI in $SiN_x$ stabilized CNFETs, from which we derive new insights into CNT interface and oxide charge trapping mechanisms. We experimentally show AC gate stress pulse cycling reduces NBTI $\Delta V_T$ accumulation, which means BEOL circuit workloads with cycled operation could have significantly, *i.e.* >$10^4$×, enhanced lifetime than what would be estimated from conservative DC NBTI measurements. This work improves our understanding of how to overcome ambient drift and BTI reliability in CNFETs, necessary for BEOL logic design and integration.

## RESULTS AND DISCUSSION
### $SiN_x$ Encapsulation Limits Ambient Drift

**Figure 1a-b** shows a schematic of our silicon BEOL compatible CNFET structure with cross-section scanning electron micrographs (SEMs) of example CNFETs integrated on 200 mm wafers in the silicon CMOS BEOL from a commercial silicon foundry.[7] Our CNFET fabrication flow [5,8]

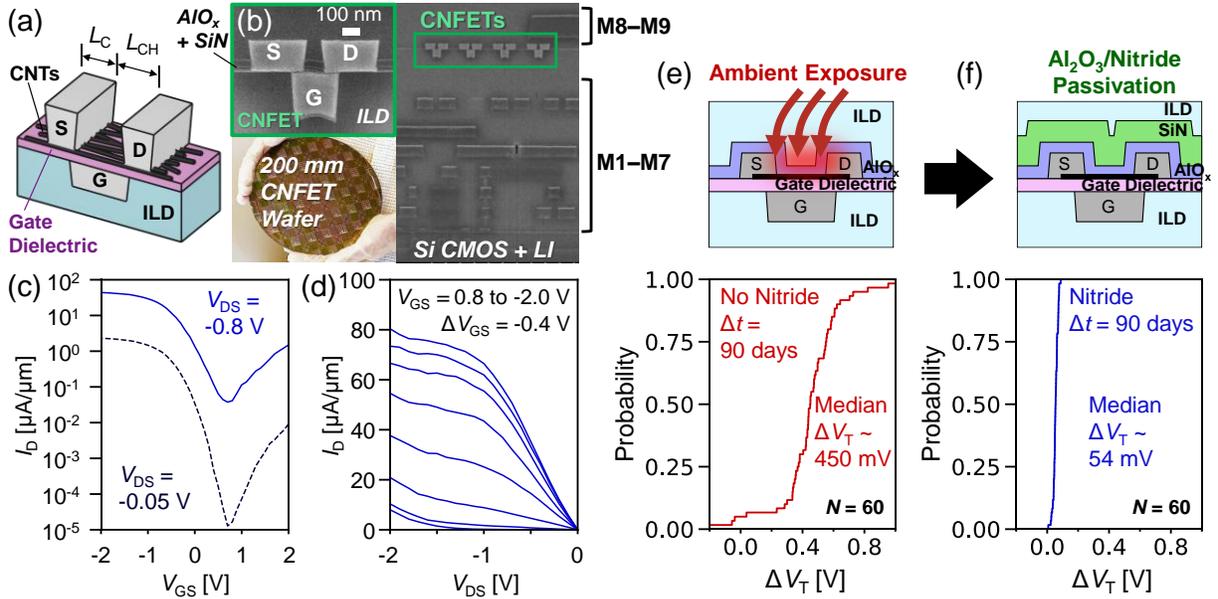

**Figure 1.** (a) Schematic of bottom gate carbon nanotube field-effect transistor (CNFET) structure integrated in foundry back-end-of-line (BEOL) with encapsulation not shown for clarity. (b) Photo of 200 mm Silicon CMOS wafer with CNFET integrated in BEOL between metal layers M7 and M8, shown in cross-section SEM micrographs. Zoom-in of CNFET with gate length $L_G$ = 260 nm, channel length $L_{CH}$ = 120 nm, contact length $L_C$ = 300 nm. Typical measured (c) $I_D$-$V_{GS}$ and (d) $I_D$-$V_{DS}$ characteristics of PMOS CNFETs ($L_{CH}$ = 120 nm, $L_C$ = 160 nm, $W$ = 4 μm, approximate linear CNT density of 20-50 CNTs/μm per [8]). CNFET silicon nitride passivation improves threshold voltage $V_T$ ambient drift over time. (e) Baseline CNFET with $Al_2O_3$ + BEOL inter-layer dielectric (ILD, $SiO_x$) encapsulation ($L_{CH}$ = 600 nm, $L_C$ = 1 μm) and CDF of extracted $\Delta V_T$ shift for each CNFET after 90 days in ambient atmosphere. (f) CNFETs with $Al_2O_3$ + $SiN_x$ + $SiO_x$ passivation and CDF of extracted $\Delta V_T$ shift for each $SiN_x$ passivated CNFET after 90 days in ambient atmosphere.

begins with a bottom gate stack made with high-k oxide ($AlO_x$ + $HfO_x$) gate dielectric deposited via atomic layer deposition (ALD) on top of embedded tungsten metal gates. After that, purified 99.99% semiconducting CNTs are deposited on top, and $O_2$ plasma is used to remove CNTs outside of the transistor channel regions. Source/drain contacts are fabricated either using evaporated metal + lift-off [6,8] or using lift-off-free etch/fill/planarization with a sputtered metal liner + tungsten fill via chemical vapor deposition (CVD).[5,7,27] **Figure 1c-d** presents typical transistor $I_D$-$V_{GS}$ and $I_D$-$V_{DS}$ electrical characteristics for lift-off-free CNFET PMOS fabricated using an ALD TiN + sputtered W metal liner which are used in this work to study NBTI. Details on transistor electrical characterization are in **Methods**.

A prior challenge in CNFETs is threshold voltage shift ($\Delta V_T$) over time from atmospheric adsorbed water doping, [12,14,16] termed "ambient drift". We find $SiN_x$ is key to passivating foundry wafer scale CNFETs from ambient atmosphere.[18] Only $Al_2O_3$ + $SiO_x$ encapsulation shows CNFET $\Delta V_T$ shift ~450 mV after 90 days exposure to atmosphere whereas $Al_2O_3$ + $SiN_x$ + $SiO_x$ encapsulation limits $\Delta V_T$ shift to ~54 mV, a ~8× improvement (**Figure 1e-f**). Additional details on $SiN_x$ encapsulation and $I_D$-$V_{GS}$ characteristics used to extract ambient $\Delta V_T$ shifts are in **Section S1 of Supporting Information**. Thus, our CNT channels in **Figure 1a-d** are encapsulated with 20 nm $Al_2O_3$ deposited via ALD at 200°C followed by plasma-enhanced chemical vapor deposition (PECVD) of 50 nm BEOL Silicon Nitride ($SiN_x$) deposited at 400°C and 200 nm BEOL dielectric ($SiO_x$) deposited at 350°C. $SiN_x$ encapsulated ambient stabilized CNFETs are used in all following sections to characterize bias temperature instability, BTI, which ensures that measured $\Delta V_T$ shift is due to BTI and not ambient drift over the course of measurement.

**DC NBTI**

A key consideration of this work is to understand the change in CNFET electrical characteristics during operation at various conditions (gate stress, temperature, frequency, etc.) to guide logic design. Negative gate bias used to operate CNFET PMOS induces negative threshold voltage shift $\Delta V_T$, i.e., negative bias temperature instability, NBTI, which is observed in silicon PMOS.[28] BTI induced $\Delta V_T$ is a key reliability challenge in CMOS circuits and must be limited to ensure correct circuit functionality over a lifetime of operation. BTI induced $\Delta V_T$ is the physical origin of hysteresis commonly observed in CNFET $I_D$-$V_{GS}$ measurements, which occurs in both unencapsulated[12,14,15] and dielectric encapsulated CNFETs.[17,18] The physical origin of NBTI in our fully dielectric encapsulated CNFET PMOS is speculated to be from oxide charge trapping (qualitative diagram in **Figure 2a**, additional discussion on our CNT-oxide interface is in **Section S2 of Supporting Information**). Filling of oxide states may be facilitated by trap assisted tunneling from interface traps at the CNT-oxide interface.[29] CNFET density of interfacial traps $D_{it}$ in this work is estimated to be $D_{it}$ ~ $10^{13}$ $cm^{-2}$ $eV^{-1}$ from $I_D$-$V_{GS}$ characteristics (details in **Section S3 of Supporting Information**), similar to prior published CNFETs[15,30] and ~1000× higher than a typical hydrogen-passivated $Si/SiO_2$ interface in conventional silicon MOSFETs.[28,31,32]

We extract NBTI $\Delta V_T$ vs. time by applying a DC constant gate stress and log-time spaced sampling the change in drain current with spot on-the-fly current measurements (**Figure 2b**, similar to conventional silicon BTI testing, with measured controller limited read pulse width ~1 ms and ~1 ms delay from stress to relaxation which is accounted for in relaxation times). Characterization of the DC NBTI measurement setup is in **Section S4 of Supporting Information**. **Figure 2c-f** shows DC NBTI threshold shift $\Delta V_T$ vs. stress and relaxation time characterized for different parameters which typically affect the rate and magnitude of NBTI [25,28]: (1) oxide electric field normal to the channel (from gate bias), normalized as $\xi_{ox}' = (V_{GS} - V_T)/EOT$ where $V_T$ here is the pre-stress threshold voltage[25,33,34], (2) temperature, and (3) equivalent oxide thickness, *EOT*. **Figure 3** shows DC NBTI $\Delta V_T$ sampled at $10^3$ second stress time vs. normalized oxide electric field sampled across multiple CNFETs. Typical NBTI models characterize $\Delta V_T$ with a power law dependence versus oxide electric field and time at a given temperature and *EOT* [25,26],

$$\Delta V_T(t) \propto \xi_{ox}'^m \cdot t^\alpha \qquad (1)$$

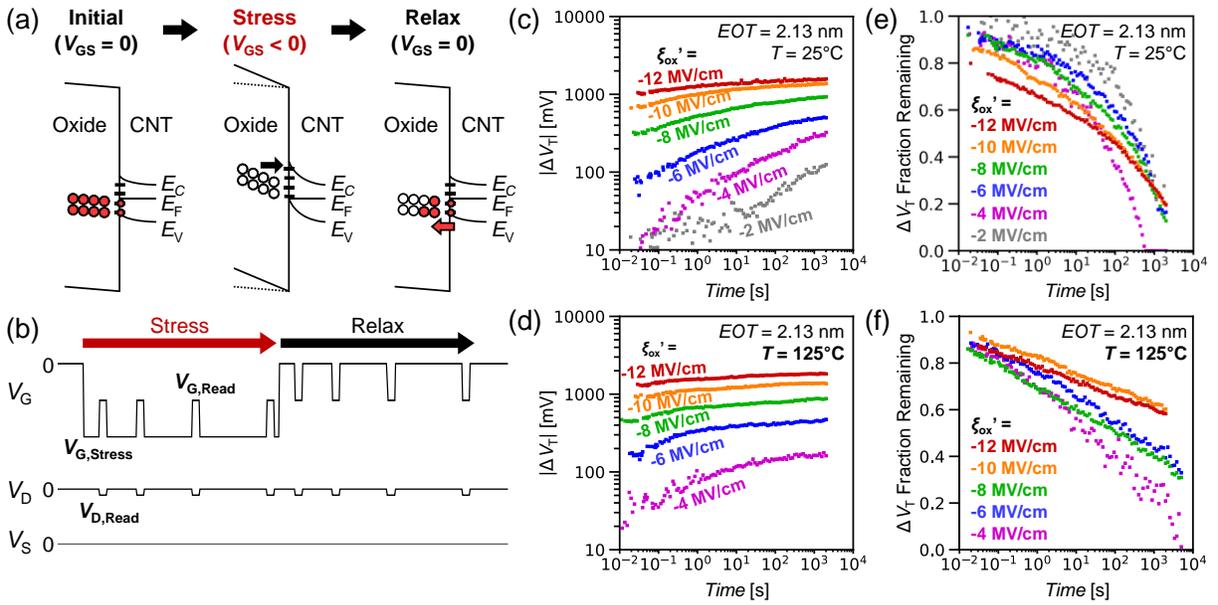

**Figure 2.** (a) Qualitative band diagrams showing speculated cause of NBTI from charge trapping and de-trapping at the CNT-oxide interface and oxide bulk traps, based on conventional silicon FET understanding.[28,29] Red circles represent electron-filled trap states and white circles represent empty trap states. (b) DC NBTI stress/relaxation on-the-fly characterization waveforms. DC NBTI in CNFET with ~70 Å gate oxide and measured $EOT \approx 2.13$ nm (effective relative dielectric constant of 12.8). (c) Room temperature and (d) 125°C DC NBTI vs. stress time at different oxide electric fields. Relative $\Delta V_T$ recovery as fraction of final stress $\Delta V_T$ vs. relaxation time at (e) room temperature and (f) 125°C. Each stress and relaxation versus time curve at each electric field value come from the same measurement.

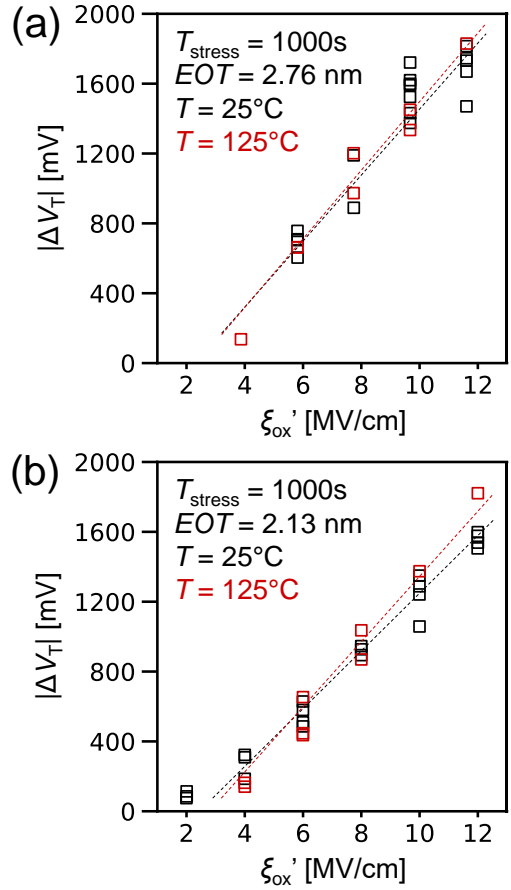

**Figure 3.** DC NBTI in CNFETs vs. gate oxide electric field $\xi_{ox}'$ at room temperature 25°C and 125°C nm sampled after $10^3$ seconds stress for (a) $EOT \approx 2.76$ nm (effective relative dielectric constant of 14.6) and (b) $EOT \approx 2.13$ nm (effective relative dielectric constant of 12.8).

where $m$ and $\alpha$ are empirical fitting parameters. From **Figure 2** and **Figure 3** we observe the following: (1) the magnitude of NBTI induced $\Delta V_T$ appears linearly dependent on normalized oxide electric field for $|\xi_{ox}'| > 2$ MV/cm; (2) temperature does not appear to increase the long-term magnitude of $\Delta V_T$ (e.g., $>10^3$ seconds); (3) however, higher temperature causes the "peak" $\Delta V_T$ to occur sooner and lengthens $\Delta V_T$ relaxation time. Additional fitting and analysis of CNFET $\Delta V_T$ relaxation with comparison to silicon FET "universal relaxation" model [35,36] is in **Section S5 of Supporting Information**.

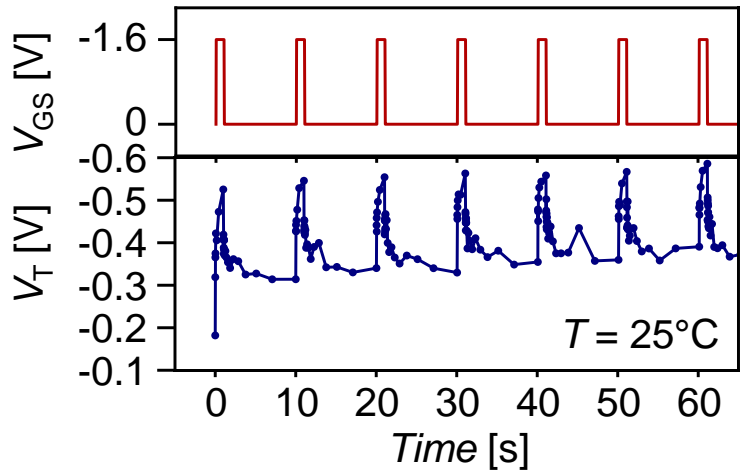

**Figure 4.** (a) Measured CNFET $V_T$ with pulsed gate stress-relax cycles, which slows buildup of BTI induced $\Delta V_T$.

**AC/Pulsed NBTI**

For typical logic applications, CNFET gate bias will not be a constant DC value, but rather will be pulsed on and off. This allows time for CNFETs to relax between bias stress, which reduces the long-term buildup of NBTI $\Delta V_T$.[37] **Figure 4** shows an example of measured CNFET $V_T$ over time with pulsed stress-relax cycles, showing how periodic relaxation cycles slow the buildup of BTI induced $\Delta V_T$. This behavior is formalized and characterized using an AC NBTI stress pattern with frequency $f$ with duty cycle $D$ which is the fraction of each cycle that the stress voltage is applied (qualitative characterization waveforms in **Figure 5a**). We apply a relax-stress-measure AC NBTI pattern which is expected to better capture both deep oxide and shallow interface traps [37]. **Figure 5b** shows reduced AC NBTI $\Delta V_T$ shift over cumulative stress time (summing only time when stress is applied, not any relaxation time) with AC stress vs. DC at 125°C across frequencies (1 kHz to 10 MHz, duty cycle $D = 50\%$). This benefit occurs even with the longer relaxation time at elevated temperature. Additional AC NBTI characterization at room temperature 25°C and with different stress-relax and relax-stress patterns[37] are in **Section S6 of Supporting Information**.

Reducing AC stress duty cycle $D$ is expected to reduce accumulated NBTI $\Delta V_T$ by providing more time for charge relaxation (**Figure 6a**). **Figure 6b** shows DC versus AC NBTI $\Delta V_T$ measured with 10 MHz stress duty cycle $D = 50\%$ and $D = 20\%$. We observe clear reduction in $\Delta V_T$ with lower AC stress duty cycle at cumulative stress times >1 second. **Figure 6c** shows the plotted trend in $\Delta V_T$ sampled at total cumulative stress time $T_{stress} = 1000$ seconds for different duty cycles and gate bias (the measured AC NBTI vs. time data for each gate bias is in **Section S7 of Supporting Information**). Our measured behavior in **Figure 6c** is fitted to a model for silicon AC NBTI based on interface trap tunneling to and from oxide states, originally developed by Tewksbury (1992),[29,37]

$$\Delta V_T(t_s, t_r) = \frac{qD_{ot}x_o}{C_{ox}}(E_F - E_{F0}) \log\left(1 + \frac{\tau_{oe} t_{stress}}{\tau_{oc} t_{relax}}\right) \quad (2)$$

$$= A \log\left(1 + B \frac{t_{stress}}{t_{relax}}\right)$$

$$\frac{t_{stress}}{t_{relax}} = \frac{D}{1-D} \quad (3)$$

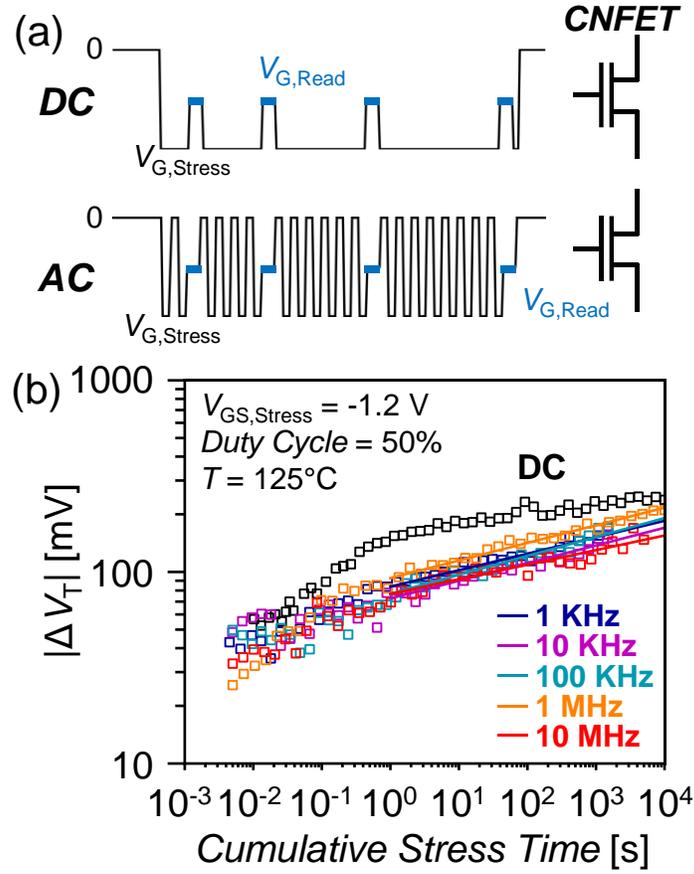

**Figure 5.** (a) DC vs. AC NBTI stress waveforms using a relax-stress-measure AC stress pattern. Controller read pulse width is 1-2 ms. (b) AC NBTI at 125°C, $EOT \approx 2.13$ nm, $V_{GS,stress} = -1.2$ V) across a range of frequency 1 kHz to 10 MHz and $D = 50\%$ duty cycle, showing lack of AC NBTI frequency dependence. Cumulative stress time sums only time when stress is applied and does not include relaxation time.

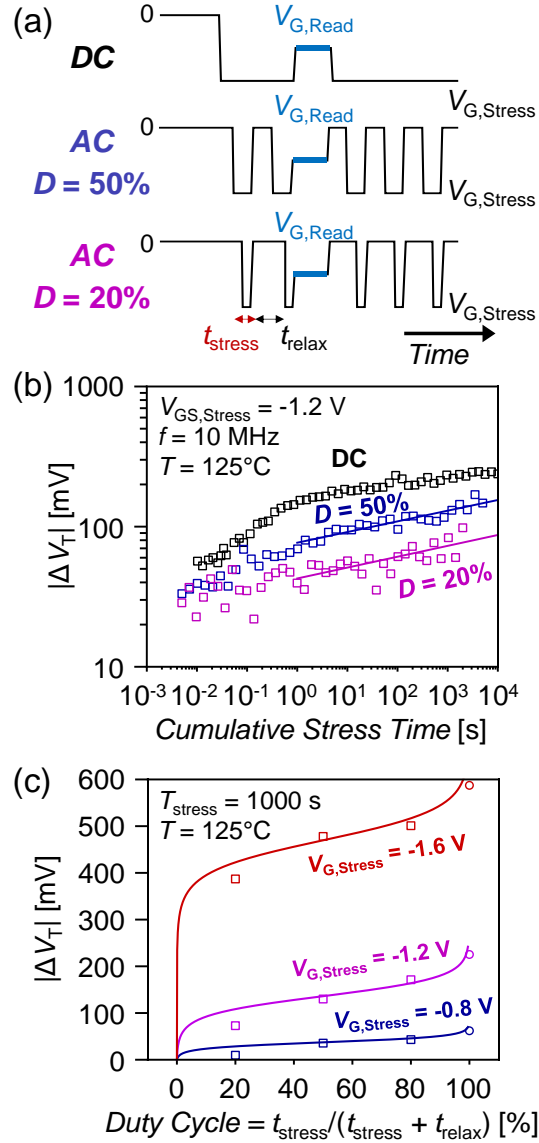

**Figure 6.** (a) DC versus pulsed AC gate voltage stress waveforms at different duty cycles *D*. (b) CNFET DC vs. AC NBTI at 125°C using 10 MHz AC stress with duty cycles *D* = 50% and D = 20% at $V_{GS,stress}$ = -1.2 V. CNFETs have $EOT \approx 2.13$ nm. (c) CNFET AC NBTI vs. duty cycle at different stress bias. Open points are measurements at 125°C, 10 MHz stress, after total accumulated stress time $T_{stress} = 10^3$ s (CNFETs have $EOT \approx 2.13$ nm). Solid lines fitted using model described in [29,37].

where $D_{ot}$ is the trap density in the oxide, $x_o$ is a characteristic tunneling depth, $C_{ox}$ is the gate dielectric capacitance, $E_F - E_{F0}$ is the energy range of traps, $\tau_{oc}$ and $\tau_{oe}$ are the trap capture and emission time constants, and $t_{stress}$ and $t_{relax}$ are the stress and relax time portions of the periodic AC stress waveform (labelled in **Figure 6a**). For model fitting, the trap dynamics are simplified into two empirical fitting parameters *A* and *B*. The general shape of this silicon charge trapping model fits the measured duty cycle dependence of $\Delta V_T$ after a fixed cumulative stress time $T_{stress}$ in our measured CNFETs undergoing AC NBTI stress. This supports the interface and oxide charge trapping mechanisms as proposed in **Figure 2a** as the origin for CNFET NBTI effects.

To project how AC duty cycling operation improves NBTI reliability, **Figure 6b** can be used to project a "NBTI time-to-failure". For example, if a circuit specification is that the max tolerable $|\Delta V_T|$ is < 100 mV, the "NBTI time-to-failure" is the cumulative stress time when the $\Delta V_T$ exceeds this threshold. Under conditions in **Figure 6b** and DC stress, this occurs at $T_{stress}$ ~ $10^{-1}$ seconds, while under AC stress with $D = 50\%$, this time-to-failure is extended out by >$10^2\times$ relative to DC, and with $D = 20\%$ it is >$10^4\times$ relative to DC. We illustrate this analysis in **Section S8 of Supporting Information**. This represents a significant NBTI tolerance enhancement under pulsed AC operation versus DC operation. This shows a methodology to operate CNFETs in pulsed logic workloads with sufficient duty cycling that can overcome NBTI.

**CONCLUSION**

In conclusion, this work advances our understanding and ability to stabilize CNFET electrical characteristics in ambient atmosphere and during operation. We experimentally demonstrate a method to limit foundry CNFET ambient atmosphere induced threshold voltage $\Delta V_T$ shift via silicon nitride encapsulation. This enables us to analyze DC and AC NBTI with negligible influence of ambient induced $\Delta V_T$. We characterize DC and AC pulsed CNFET NBTI at various operating conditions (gate stress, temperature, frequency, duty cycle). We show DC NBTI is conservative and AC operation significantly limits NBTI induced $\Delta V_T$ accumulation at a wide range of gate voltage bias and AC stress frequency. These results and techniques enable future work to develop more accurate models for circuit designers to co-design appropriate circuit-level tolerances and mitigations for CNFET behavior shift.

**METHODS**

CNFETs in this work are fabricated in a silicon CMOS foundry on 200 mm wafers using conventional front-end-of-line silicon tooling. Details on CNFET process flow are described in [5,7,8]. CNFET transistor $I_D$-$V_{GS}$ and $I_D$-$V_{DS}$ electrical measurements are performed in air at room temperature in a Cascade Summit 12000 probe station with a Keysight B1500A semiconductor parameter analyzer. Room temperature BTI stress characterization is performed in a Cascade

Summit 12000 probe station in atmosphere ambient conditions. Elevated temperature stress is performed in a Cascade probe station with a heated chuck in atmosphere with nitrogen flowing over the wafer. Pulsed on-the-fly BTI measurements are implemented using a National Instruments chassis with a PXIE 6570 pulse driver interfaced using python control scripts. Characterization of the DC and AC NBTI measurement setup is in **Section S4, S6 of Supporting Information**.

## AUTHOR INFORMATION

**Corresponding Author**

**Tathagata Srimani** – *Department of Electrical and Computer Engineering, Carnegie Mellon University, Pittsburgh, PA 15213, USA; Department of Electrical Engineering, Stanford University, Stanford, CA 94305, USA;* orcid.org/0000-0002-1238-7324; *Email:* tsrimani@andrew.cmu.edu

**Authors**

**Andrew C. Yu** – *Department of Electrical Engineering and Computer Science, Massachusetts Institute of Technology, Cambridge, MA 02139, USA;* orcid.org/0000-0002-4596-9101

**Max M. Shulake**r – *Department of Electrical Engineering and Computer Science, Massachusetts Institute of Technology, Cambridge, MA 02139, USA; Analog Devices, Wilmington, MA 01887, USA;* orcid.org/0000-0003-2237-193X

## ASSOCIATED CONTENT

**Supporting Information**.

The Supporting Information contains:

> CNFET electrical measurements with/without silicon nitride (Section S1), CNT-oxide interface (Section S2), CNFET density of interfacial traps estimation (Section S3), DC NBTI setup characterization (Section S4), CNFET relaxation (Section S5), CNFET AC NBTI setup and additional characterization (Section S6), CNFET AC NBTI additional duty cycle characterization (Section S7), and illustration of AC NBTI time-to-failure improvement vs. DC (Section S8).

## AUTHOR INFORMATION

**Corresponding Author**

Tathagata Srimani. Email: tsrimani@andrew.cmu.edu

**Author Contributions**

The manuscript was written through contributions of all authors. All authors have given approval to the final version of the manuscript.

**Funding Sources**

We acknowledge support from the DARPA 3DSoC program, Analog Devices, NSF Graduate Fellowship, U.S. DoD Eccalon Project, and NSF FuSe2 Award 2425218. The views, opinions, and/or findings expressed are those of the authors and should not be interpreted as representing the official views or policies of the Department of Defense or the US Government.

## ACKNOWLEDGMENT

We thank Christian Lau from MIT, and Dennis Rich, Gilad Zeevi, Andrew Bechdolt, and Subhasish Mitra from Stanford University for idea refinement/discussions.

## REFERENCES

(1) Srimani, T.; Radway, R. M.; Kim, J.; Prabhu, K.; Rich, D.; Gilardi, C.; Raina, P.; Shulaker, M.; Lim, S. K.; Mitra, S. Ultra-Dense 3D Physical Design Unlocks New Architectural Design Points with Large Benefits. In *2023 Design, Automation & Test in Europe Conference & Exhibition (DATE)*; IEEE: Antwerp, Belgium, 2023; pp 1–6. https://doi.org/10.23919/DATE56975.2023.10137051.

(2) Srimani, T.; Bechdolt, A.; Choi, S.; Gilardi, C.; Kasperovich, A.; Li, S.; Lin, Q.; Malakoutian, M.; McEwen, P.; Radway, R. M.; Rich, D.; Yu, A. C.; Fuller, S.; Achour, S.; Chowdhury, S.; Wong, H.-S. P.; Shulaker, M.; Mitra, S. N3XT 3D Technology Foundations and Their Lab-to-Fab: Omni 3D Logic, Logic+Memory Ultra-Dense 3D, 3D Thermal Scaffolding. In *2023 International Electron Devices Meeting (IEDM)*; IEEE: San Francisco, CA, USA, 2023; pp 1–4. https://doi.org/10.1109/IEDM45741.2023.10413794.

(3) Srimani, T.; Ding, J.; Yu, A.; Kanhaiya, P.; Lau, C.; Ho, R.; Humes, J.; Kingston, C. T.; Malenfant, P. R. L.; Shulaker, M. M. Comprehensive Study on High Purity Semiconducting Carbon Nanotube Extraction. *Adv Elect Materials* **2022**, *8* (9), 2101377. https://doi.org/10.1002/aelm.202101377.

(4) Lau, C.; Srimani, T.; Bishop, M. D.; Hills, G.; Shulaker, M. M. Tunable *n*-Type Doping of Carbon Nanotubes through Engineered Atomic Layer Deposition HfO$_X$ Films. *ACS Nano* **2018**, *12* (11), 10924–10931. https://doi.org/10.1021/acsnano.8b04208.

(5) Yu, A. C.; Srimani, T.; Lau, C.; Benton, B.; Nelson, M.; Shulaker, M. M. Foundry Integration of Carbon Nanotube FETs With 320 Nm Contacted Gate Pitch Using New Lift-Off-Free Process. *IEEE Electron Device Lett.* **2022**, *43* (3), 486–489. https://doi.org/10.1109/LED.2022.3144936.

(6) Srimani, T.; Hills, G.; Bishop, M.; Lau, C.; Kanhaiya, P.; Ho, R.; Amer, A.; Chao, M.; Yu, A.; Wright, A.; Ratkovich, A.; Aguilar, D.; Bramer, A.; Cecman, C.; Chov, A.; Clark, G.; Michaelson, G.; Johnson, M.; Kelley, K.; Manos, P.; Mi, K.; Suriono, U.; Vuntangboon, S.; Xue, H.; Humes, J.; Soares, S.; Jones, B.; Burack, S.; Arvind; Chandrakasan, A.; Ferguson, B.; Nelson, M.; Shulaker, M. M. Heterogeneous Integration of BEOL Logic and Memory in a Commercial Foundry: Multi-Tier Complementary Carbon Nanotube Logic and Resistive RAM at a 130 Nm Node. In *2020 IEEE Symposium on VLSI Technology*; IEEE: Honolulu, HI, USA, 2020; pp 1–2. https://doi.org/10.1109/VLSITechnology18217.2020.9265083.

(7) Srimani, T.; Yu, A. C.; Radway, R. M.; Rich, D. T.; Nelson, M.; Wong, S.; Murphy, D.; Fuller, S.; Hills, G.; Mitra, S.; Shulaker, M. M. Foundry Monolithic 3D BEOL Transistor + Memory Stack: Iso-Performance and Iso-Footprint BEOL Carbon Nanotube FET+RRAM vs. FEOL Silicon FET+RRAM. *2023 IEEE Symposium on VLSI Technology and Circuits (VLSI Technology and Circuits)* **2023**, 1–2. https://doi.org/10.23919/VLSITechnologyandCir57934.2023.10185414.

(8) Bishop, M. D.; Hills, G.; Srimani, T.; Lau, C.; Murphy, D.; Fuller, S.; Humes, J.; Ratkovich, A.; Nelson, M.; Shulaker, M. M. Fabrication of Carbon Nanotube Field-Effect Transistors in Commercial Silicon Manufacturing Facilities. *Nat Electron* **2020**, *3* (8), 492–501. https://doi.org/10.1038/s41928-020-0419-7.

(9) Hills, G.; Lau, C.; Wright, A.; Fuller, S.; Bishop, M. D.; Srimani, T.; Kanhaiya, P.; Ho, R.; Amer, A.; Stein, Y.; Murphy, D.; Arvind; Chandrakasan, A.; Shulaker, M. M. Modern Microprocessor Built from Complementary Carbon Nanotube Transistors. *Nature* **2019**, *572* (7771), 595–602. https://doi.org/10.1038/s41586-019-1493-8.

(10) Kanhaiya, P. S.; Lau, C.; Hills, G.; Bishop, M. D.; Shulaker, M. M. Carbon Nanotube-Based CMOS SRAM: 1 Kbit 6T SRAM Arrays and 10T SRAM Cells. *IEEE Trans. Electron Devices* **2019**, *66* (12), 5375–5380. https://doi.org/10.1109/TED.2019.2945533.

(11) Hills, G.; Bardon, M. G.; Doornbos, G.; Yakimets, D.; Schuddinck, P.; Baert, R.; Jang, D.; Mattii, L.; Sherazi, S. M. Y.; Rodopoulos, D.; Ritzenthaler, R.; Lee, C.-S.; Thean, A. V.-Y.; Radu, I.; Spessot, A.; Debacker, P.; Catthoor, F.; Raghavan, P.; Shulaker, M. M.; Wong, H.-S. P.; Mitra, S. Understanding Energy Efficiency Benefits of Carbon Nanotube Field-Effect Transistors for Digital VLSI. *IEEE Trans. Nanotechnology* **2018**, *17* (6), 1259–1269. https://doi.org/10.1109/TNANO.2018.2871841.

(12) Kim, W.; Javey, A.; Vermesh, O.; Wang, Q.; Li, Y.; Dai, H. Hysteresis Caused by Water Molecules in Carbon Nanotube Field-Effect Transistors. *Nano Lett.* **2003**, *3* (2), 193–198. https://doi.org/10.1021/nl0259232.

(13) Lee, J. S.; Ryu, S.; Yoo, K.; Choi, I. S.; Yun, W. S.; Kim, J. Origin of Gate Hysteresis in Carbon Nanotube Field-Effect Transistors. *J. Phys. Chem. C* **2007**, *111* (34), 12504–12507. https://doi.org/10.1021/jp074692q.

(14) Estrada, D.; Dutta, S.; Liao, A.; Pop, E. Reduction of Hysteresis for Carbon Nanotube Mobility Measurements Using Pulsed Characterization. *Nanotechnology* **2010**, *21* (8), 085702. https://doi.org/10.1088/0957-4484/21/8/085702.

(15) Park, R. S.; Shulaker, M. M.; Hills, G.; Suriyasena Liyanage, L.; Lee, S.; Tang, A.; Mitra, S.; Wong, H.-S. P. Hysteresis in Carbon Nanotube Transistors: Measurement and Analysis of Trap Density, Energy Level, and Spatial Distribution. *ACS Nano* **2016**, *10* (4), 4599–4608. https://doi.org/10.1021/acsnano.6b00792.

(16) Park, R. S.; Hills, G.; Sohn, J.; Mitra, S.; Shulaker, M. M.; Wong, H.-S. P. Hysteresis-Free Carbon Nanotube Field-Effect Transistors. *ACS Nano* **2017**, *11* (5), 4785–4791. https://doi.org/10.1021/acsnano.7b01164.

(17) Zhao, Y.; Huo, Y.; Xiao, X.; Wang, Y.; Zhang, T.; Jiang, K.; Wang, J.; Fan, S.; Li, Q. Inverse Hysteresis and Ultrasmall Hysteresis Thin-Film Transistors Fabricated Using Sputtered Dielectrics. *Adv Elect Materials* **2017**, *3* (3), 1600483. https://doi.org/10.1002/aelm.201600483.

(18) Lau, C. A Manufacturing Methodology for Carbon Nanotube-Based Digital Systems: From Devices, to Doping, to System Demonstrations, Massachusetts Institute of Technology, 2022. https://dspace.mit.edu/handle/1721.1/143373.

(19) Wang, Y.; Wang, S.; Ye, H.; Zhang, W.; Xiang, L. Negative Bias Temperature Instability in Top-Gated Carbon Nanotube Thin Film Transistors With $Y_2O_3$ Gate Dielectric. *IEEE Trans. Device Mater. Relib.* **2023**, *23* (4), 571–576. https://doi.org/10.1109/TDMR.2023.3322157.

(20) Noyce, S. G.; Doherty, J. L.; Cheng, Z.; Han, H.; Bowen, S.; Franklin, A. D. Electronic Stability of Carbon Nanotube Transistors Under Long-Term Bias Stress. *Nano Lett.* **2019**, *19* (3), 1460–1466. https://doi.org/10.1021/acs.nanolett.8b03986.

(21) Sun, Y.; Lu, P.; Zhang, L.; Cao, Y.; Bai, L.; Ding, L.; Han, J.; Zhang, C.; Zhu, M.; Zhang, Z. Investigation and Improvement of the Bias Temperature Instability in Carbon Nanotube Transistors. *Adv Elect Materials* **2024**, 2400464. https://doi.org/10.1002/aelm.202400464.

(22) Reisinger, H.; Brunner, U.; Heinrigs, W.; Gustin, W.; Schlunder, C. A Comparison of Fast Methods for Measuring NBTI Degradation. *IEEE Trans. Device Mater. Relib.* **2007**, *7* (4), 531–539. https://doi.org/10.1109/TDMR.2007.911385.

(23) Ming-Fu Li; Daming Huang; Chen Shen; Yang, T.; Liu, W. J.; Zhiying Liu. Understand NBTI Mechanism by Developing Novel Measurement Techniques. *IEEE Trans. Device Mater. Relib.* **2008**, *8* (1), 62–71. https://doi.org/10.1109/TDMR.2007.912273.

(24) Grasser, T.; Wagner, P.-Jü.; Hehenberger, P.; Goes, W.; Kaczer, B. A Rigorous Study of Measurement Techniques for Negative Bias Temperature Instability. *IEEE Trans. Device Mater. Relib.* **2008**, *8* (3), 526–535. https://doi.org/10.1109/TDMR.2008.2002353.

(25) Cho, M.; Lee, J.-D.; Aoulaiche, M.; Kaczer, B.; Roussel, P.; Kauerauf, T.; Degraeve, R.; Franco, J.; Ragnarsson, L.-Å.; Groeseneken, G. Insight Into N/PBTI Mechanisms in Sub-1-Nm-EOT Devices. *IEEE Trans. Electron Devices* **2012**, *59* (8), 2042–2048. https://doi.org/10.1109/TED.2012.2199496.

(26) Kerber, A.; Cartier, E. A. Reliability Challenges for CMOS Technology Qualifications With Hafnium Oxide/Titanium Nitride Gate Stacks. *IEEE Trans. Device Mater. Relib.* **2009**, *9* (2), 147–162. https://doi.org/10.1109/TDMR.2009.2016954.

(27) Srimani, T.; Yu, A. C.; Benton, B.; Nelson, M.; Shulaker, M. M. Lift-off-Free Complementary Carbon Nanotube FETs Fabricated With Conventional Processing in a Silicon Foundry. In *2022 International Symposium on VLSI Technology, Systems and*

*Applications (VLSI-TSA)*; IEEE: Hsinchu, Taiwan, 2022; pp 1–2. https://doi.org/10.1109/VLSI-TSA54299.2022.9771013.

(28) Schroder, D. K. Negative Bias Temperature Instability: What Do We Understand? *Microelectronics Reliability* **2007**, *47* (6), 841–852. https://doi.org/10.1016/j.microrel.2006.10.006.

(29) Tewksbury, T. L. Relaxation Effects in MOS Devices Due to Tunnel Exchange with Near-Interface Oxide Traps, Massachusetts Institute of Technology, 1992. https://dspace.mit.edu/handle/1721.1/13238.

(30) Li, S.; Chao, T.-A.; Gilardi, C.; Safron, N.; Su, S.-K.; Zeevi, G.; Bechdolt, A. D.; Passlack, M.; Oberoi, A.; Lin, Q.; Zhang, Z.; Wang, K.; Kashyap, H.; Liew, S.-L.; Hou, V. D.-H.; Kummel, A.; Radu, L.; Pitner, G.; Wong, H.-S. P.; Mitra, S. High-Performance and Low Parasitic Capacitance CNT MOSFET: 1.2 mA/Mm at $V_{DS}$ of 0.75 V by Self-Aligned Doping in Sub-20 Nm Spacer. In *2023 International Electron Devices Meeting (IEDM)*; IEEE: San Francisco, CA, USA, 2023; pp 1–4. https://doi.org/10.1109/IEDM45741.2023.10413827.

(31) Sah Chih-Tang. Evolution of the MOS Transistor-from Conception to VLSI. *Proc. IEEE* **1988**, *76* (10), 1280–1326. https://doi.org/10.1109/5.16328.

(32) Castro, P. L.; Deal, B. E. Low-Temperature Reduction of Fast Surface States Associated with Thermally Oxidized Silicon. *J. Electrochem. Soc.* **1971**, *118* (2), 280. https://doi.org/10.1149/1.2408016.

(33) Mukhopadhyay, S.; Joshi, K.; Chaudhary, V.; Goel, N.; De, S.; Pandey, R. K.; Murali, K. V. R. M.; Mahapatra, S. Trap Generation in IL and HK Layers during BTI / TDDB Stress in Scaled HKMG N and P MOSFETs. In *2014 IEEE International Reliability Physics Symposium*; IEEE: Waikoloa, HI, 2014; p GD.3.1-GD.3.11. https://doi.org/10.1109/IRPS.2014.6861146.

(34) Degraeve, R.; Aoulaiche, M.; Kaczer, B.; Roussel, Ph.; Kauerauf, T.; Sahhaf, S.; Groeseneken, G. Review of Reliability Issues in High-k/Metal Gate Stacks. In *2008 15th International Symposium on the Physical and Failure Analysis of Integrated Circuits*; IEEE: singapore, 2008; pp 1–6. https://doi.org/10.1109/IPFA.2008.4588195.

(35) Kaczer, B.; Grasser, T.; Roussel, J.; Martin-Martinez, J.; O'Connor, R.; O'Sullivan, B. J.; Groeseneken, G. Ubiquitous Relaxation in BTI Stressing—New Evaluation and Insights. In *2008 IEEE International Reliability Physics Symposium*; IEEE: Phoenix, AZ, 2008; pp 20–27. https://doi.org/10.1109/RELPHY.2008.4558858.

(36) Grasser, T.; Gös, W.; Sverdlov, V.; Kaczer, B. The Universality of NBTI Relaxation and Its Implications for Modeling and Characterization. In *2007 IEEE International Reliability Physics Symposium Proceedings. 45th Annual*; IEEE: Phoenix, AZ, USA, 2007; pp 268–280. https://doi.org/10.1109/RELPHY.2007.369904.

(37) Tsai, Y. S.; Jha, N. K.; Lee, Y.-H.; Ranjan, R.; Wang, W.; Shih, J. R.; Chen, M. J.; Lee, J. H.; Wu, K. Prediction of NBTI Degradation for Circuit under AC Operation. In *2010 IEEE International Reliability Physics Symposium*; IEEE: Garden Grove (Anaheim), CA, USA, 2010; pp 665–669. https://doi.org/10.1109/IRPS.2010.5488752.